\documentclass[preprint,review,12pt]{elsarticle}

\usepackage{amssymb}
\usepackage{subfigure}
\usepackage{amsmath}
\usepackage[utf8]{inputenc}

\journal{Optics Communications}

\newcommand{\pra}{Phys. Rev. A\ }
\newcommand{\prl}{Phys. Rev. Lett.\ }
\newcommand{\pr}{Phys. Rev.\ }

\newcommand{\jpa}{J. Phys. A\ }
\newcommand{\jpb}{J. Phys. B\ }
\newcommand{\apl}{App. Phys. Lett.\ }
\newcommand{\ol}{Opt. Lett.\ }
\newcommand{\etal}{{\em et al.}\ }
\newcommand{\etals}{{\em et al.}}

\newcommand{\oc}{Opt. Commun.\ }
\newcommand{\josab}{J. Opt. Soc. Am. B\ }

\begin{document}

\begin{frontmatter}

\title{Tripartite correlations over two octaves from cascaded harmonic generation}

\author[mko]{M.K.~Olsen}
\ead{mko@physics.uq.edu.au}

\address{School of Mathematics and Physics, University of Queensland, Brisbane, 
Queensland 4072, Australia.}

\begin{abstract}

We analyse the output quantum tripartite correlations from an intracavity nonlinear optical system which uses cascaded nonlinearities to produce both second and fourth harmonic outputs from an input field at the fundamental frequency. Using fully quantum equations of motion, we investigate two parameter regimes and show that  the system produces 
tripartite inseparability, entanglement and EPR steering, with the detection of these depending on the correlations being considered.

\end{abstract}


\end{frontmatter}

\section{Introduction}
\label{sec:intro}

Fourth harmonic generation has not received a huge amount of attention in the scientific literature, possibly because materials with the nonlinearity needed for a five wave mixing process are difficult to find. Despite this inherent problem, and the difficulty of finding materials that are transparent over two octaves, Komatsu \etal have successfully produced fourth harmonic from Li$_{2}$B$_{4}$O$_{7}$ crystal, with a conversion efficiency of $20\%$~\cite{Kumatsu}. The advent of quasi-periodic superlattices meant that higher than second order processes were now available, with Zhu \etal producing third harmonic by coupling second harmonic (SHG) and sum-frequency generation in 1997~\cite{Zhu}. Using CsLiB$_{6}$O$_{10}$, Kojima \etal were able to produce fourth harmonic at a 10 kHz repetition rate by 2000~\cite{Kojima}. Broderick \etal have produced fourth harmonic from a cascaded SHG process using a  HeXLN crystal~\cite{Broderick} tuneable for both processes at the same temperature. Südmeyer \etal produced fields at both second and fourth harmonics using an intracavity cascaded process with LBO and BBO crystals, with greater than $50\%$ efficiency in 2007~\cite{Sudmeyer}. More recently, Ji \etal have generated light at 263 nm from a 1053 nm input, using KD$^{*}$P and NH$_{4}$H$_{2}$PO$_{4}$ crystals with non-critical phase matching~\cite{Ji}.

The theoretical examination of the quantum statistical properties of fourth harmonic generation began with Kheruntsyan \etals, who analysed an intracavity cascaded frequency doubler process~\cite{Yerevan}. The authors adiabatically eliminated the highest frequency mode to calculate squeezing in the lower modes, also finding self-pulsing in the intensities. Yu and Wang~\cite{Yu} performed an analysis of the system without any elimination, starting with the full positive-P representation~\cite{P+} equations of motion. Linearising around the steady-state solutions of the semi-classical equations, they performed a stability analysis and examined the entanglement properties using the method of symplectic eigenvalues~\cite {Serafini}. More recently, Olsen has examined the quantum statistical properties of the system~\cite{4HGbi}, finding that quadrature squeezing and bipartite entanglement and asymmetric Einstein-Podolsky-Rosen (EPR) steering~\cite{SFG,sapatona} are available for some of the possible bipartitions.  

In this work we extend previous analyses by examining various correlations often used in continuous variable systems to detect tripartite inseparability, entanglement, and steering. We begin with the two types of van Loock-Furusawa (vLF) inequalities~\cite{vLF} and their refinements by Teh and Reid~\cite{Teh} for mixed states. Following from that, we will use three mode EPR~\cite{EPR} inequalities developed by Olsen, Bradley and Reid (OBR)~\cite{EPR3}, to investigate whether two members of a possible tripartition can combine to steer the third and whether our results can indicate genuine multipartite steering~\cite{HeReid}. We investigate two different regimes with changes in the pumping rate, the loss rates, and the ratio of the two $\chi^{(2)}$ nonlinearities.

\section{Hamiltonian and equations of motion}
\label{sec:model}

The system consists of three optical fields interacting in nonlinear media, which could either be a periodically poled dielectric or two separate nonlinear crystals held in the same optical cavity. The equations of motion are the same for both. The fundamental field at $\omega_{1}$, which will be externally pumped, is represented by $\hat{a}_{1}$. The second harmonic, at $\omega_{2}=2\omega_{1}$, is represented by $\hat{a}_{2}$, and the fourth harmonic, at $\omega_{3}=4\omega_{1}$, is represented by $\hat{a}_{3}$. The nonlinearity $\kappa_{1}$ couples the fields at $\omega_{1}$ and $\omega_{2}$, while $\kappa_{2}$ couples those at $\omega_{2}$ and $\omega_{3}$. The unitary interaction Hamiltonian in a rotating frame is then written as 
\begin{equation}
{\cal H}_{int} = \frac{i\hbar}{2}\left[ \kappa_{1}(\hat{a}_{1}^{2}\hat{a}_{2}^{\dag}-\hat{a}_{1}^{\dag\,2}\hat{a}_{2})+\kappa_{2}(\hat{a}_{2}^{2}\hat{a}_{3}^{\dag}-\hat{a}_{2}^{\dag\,2}\hat{a}_{3}) \right].
\label{eq:UHam}
\end{equation}
The cavity pumping Hamiltonian is
\begin{equation}
{\cal H}_{pump} = i\hbar\left(\epsilon\hat{a}_{1}^{\dag}-\epsilon^{\ast}\hat{a}_{1} \right),
\label{eq:Hpump}
\end{equation}
where $\epsilon$ represents an external pumping field which is usually taken as coherent, although this is not necessary~\cite{Liz}. The damping of the cavity into a zero temperature Markovian reservoir is described by the Lindblad superoperator 
\begin{equation}
{\cal L}\rho = \sum_{i=1}^{3}\gamma_{i}\left(2\hat{a}_{i}\rho\hat{a}_{i}^{\dag}-\hat{a}_{i}^{\dag}\hat{a}_{i}\rho-\rho\hat{a}_{i}^{\dag}\hat{a}_{i} \right),
\label{eq:Lindblad}
\end{equation}
where $\rho$ is the system density matrix and $\gamma_{i}$ is the cavity loss rate at $\omega_{i}$. We will treat all three fields as being at resonance with the optical cavity, which means that we do not need to examine quadrature correlations at all angles, but that the canonical $\hat{X}$ and $\hat{Y}$ quadratures are sufficient.

Following the usual procedures~\cite{DFW,QNoise}, we proceed via the von Neumann and Fokker-Planck equations to derive equations of motion in the positive-P representation~\cite{P+},
\begin{eqnarray}
\frac{d\alpha_{1}}{dt} &=& \epsilon-\gamma_{1}\alpha_{1}+\kappa_{1}\alpha_{1}^{+}\alpha_{2}+\sqrt{\kappa_{1}\alpha_{2}}\,\eta_{1}, \nonumber \\
\frac{d\alpha_{1}^{+}}{dt} &=& \epsilon^{\ast}-\gamma_{1}\alpha_{1}^{+}+\kappa_{1}\alpha_{1}\alpha_{2}^{+}+\sqrt{\kappa_{1}\alpha_{2}^{+}}\,\eta_{2}, \nonumber \\
\frac{d\alpha_{2}}{dt} &=& -\gamma_{2}\alpha_{2}+\kappa_{2}\alpha_{2}^{+}\alpha_{3}-\frac{\kappa_{1}}{2}\alpha_{1}^{2}+\sqrt{\kappa_{2}\alpha_{3}}\,\eta_{3}, \nonumber \\
\frac{d\alpha_{2}^{+}}{dt} &=& -\gamma_{2}\alpha_{2}^{+}+\kappa_{2}\alpha_{2}\alpha_{3}^{+}-\frac{\kappa_{1}}{2}\alpha_{1}^{+\,2}+\sqrt{\kappa_{2}\alpha_{3}^{+}}\,\eta_{4}, \nonumber \\
\frac{d\alpha_{3}}{dt} &=& -\gamma_{3}\alpha_{3}-\frac{\kappa_{2}}{2}\alpha_{2}^{2}, \nonumber \\
\frac{d\alpha_{3}^{+}}{dt} &=& -\gamma_{3}\alpha_{3}^{+}-\frac{\kappa_{2}}{2}\alpha_{2}^{+\,2},
\label{eq:Pplus}
\end{eqnarray}
noting that these have the same form in either It\^o or Stratonovich calculus~\cite{SMCrispin}.
The complex variable pairs $(\alpha_{i},\alpha_{j}^{+})$ correspond to the operator pairs $(\hat{a}_{i},\hat{a}_{j}^{\dag})$ in the sense that stochastic averages of products converge to normally-ordered operator expectation values, e.g. $\overline{\alpha_{i}^{+\,m}\alpha_{j}^{n}} \rightarrow \langle \hat{a}_{i}^{\dag\,m}\hat{a}_{j}^{n} \rangle$. The $\eta_{j}$ are Gaussian noise terms with the properties $\overline{\eta_{i}}=0$ and $\overline{\eta_{j}(t)\eta_{k}(t')}=\delta_{jk}\delta(t-t')$.

\section{Quantum correlations}
\label{sec:interest}      

Before defining the inequalities we will use, we define the amplitude quadratures of the three interacting fields as
\begin{eqnarray}
\hat{X}_{i} &=& \hat{a}_{i}+\hat{a}_{i}^{\dag}, \nonumber \\
\hat{Y}_{i} &=& -i\left(\hat{a}_{i}-\hat{a}_{i}^{\dag} \right),
\label{eq:quads}
\end{eqnarray}
with the Heisenberg uncertainty principal demanding that the product of the variances, $V(\hat{X}_{i})V(\hat{Y}_{i})\geq 1$.

For three mode inseparability and entanglement, we use the van-Loock Furusawa inequalities~\cite{vLF}, which have proven useful for other cascaded optical systems~\cite{AxMuzzJPB}. The first of these is 
\begin{equation}
V_{ij} = V(\hat{X}_{i}-\hat{X}_{j})+V(\hat{Y}_{i}+\hat{Y}_{j}+g_{k}\hat{Y}_{k}) \geq 4, 
\label{eq:VLF}
\end{equation}
for which the violation of any two demonstrates tripartite inseparability. The $g_{j}$, which are arbitrary and real, can be optimised~\cite{AxMuzz}, using the variances and covariances, as
\begin{equation}
g_{i} = -\frac{V(\hat{Y}_{i},\hat{Y}_{j})+V(\hat{Y}_{i},\hat{Y}_{k})}{V(\hat{Y}_{i})}.
\label{eq:VLFopt}
\end{equation}
Teh and Reid~\cite{Teh} have shown that, for mixed states, tripartite entanglement is demonstrated if the sum of the three correlations is less than $8$, with genuine tripartite EPR (Einstein-Podolsky-Rosen)-steering~\cite{EPR,Erwin,Wisesteer} requiring a sum of less than $4$. 

The second set set of vLF inequalities,
\begin{equation}
V_{ijk} = V(\hat{X}_{i}-\frac{\hat{X}_{j}+\hat{X}_{k}}{\sqrt{2}})+V(\hat{Y}_{i}+\frac{\hat{Y}_{j}+\hat{Y}_{k}}{\sqrt{2}}) \geq 4,
\label{eq:VLFijk}
\end{equation}
requires the violation of only one to prove tripartite inseparability. Teh and Reid~\cite{Teh} also showed that for mixed states any one of these less than $2$ demonstrates genuine tripartite entanglement, while one of them less than $1$ demonstrates genuine tripartite EPR steering. Because our nonlinear system is held in a cavity which is open to the environment, we are working with mixed states here. 

For multipartite EPR-steering, Wang \etal  showed that the steering of a given quantum mode is allowed when not less than half of the total number of modes take part in the steering group~\cite{halfplus}. In a tripartite system, this means that measurements on two of the modes are needed to steer the third. In order to quantify this, we will use the correlation functions developed by Olsen, Bradley, and Reid~\cite{EPR3}. With tripartite inferred variances as
\begin{eqnarray}
V_{inf}^{(t)}(\hat{X}_{i}) &=& V(\hat{X}_{i})-\frac{\left[V(\hat{X}_{i},\hat{X}_{j}\pm\hat{X}_{k})\right]^{2}}{V(\hat{X}_{j}\pm\hat{X}_{k})}, \nonumber \\
V_{inf}^{(t)}(\hat{Y}_{i}) &=& V(\hat{Y}_{i})-\frac{\left[V(\hat{Y}_{i},\hat{Y}_{j}\pm\hat{Y}_{k})\right]^{2}}{V(\hat{Y}_{j}\pm\hat{Y}_{k})}, 
\label{eq:V3inf}
\end{eqnarray}
we define
\begin{equation}
OBR_{ijk} = V_{inf}^{(t)}(\hat{X}_{i})V_{inf}^{(t)}(\hat{Y}_{i}),
\label{eq:OBR}
\end{equation}
so that a value of less than one means that mode $i$ can be steered by the combined forces of modes $j$ and $k$. According to the work of He and Reid~\cite{HeReid}, genuine tripartite steering is demonstrated whenever
\begin{equation}
OBR_{ijk}+OBR_{jki}+OBR_{kij} < 1.
\label{eq:genuinetristeer}
\end{equation}
In this work we will use only the plus signs in Eq.~\ref{eq:V3inf}, which will be denoted on the figure axes as OBR$_{ijk}^{+}$. We found that this gave greater violations of the inequalities in some cases, although the results were not qualitatively different.

\section{Steady-state spectral correlations}
\label{sec:cavidade}

We find that the semi-classical and quantum solutions for the intensities are identical until a certain pump power, after which the system enters a self-pulsing regime~\cite{Yerevan,4HGbi,SHGpulse,THGcascade}. Below this pump power, the steady-state solutions for the field amplitudes found from the integration of the full positive-P equations and their semiclassical equivalents  are identical. The semiclassical equations are found by removing the noise terms from Eq.~\ref{eq:Pplus}, and have been solved numerically here. 

The measured observables of an intracavity process are usually the output spectral correlations, which are accessible using homodyne measurement techniques~\cite{mjc}. These are readily calculated in the steady-state by treating the system as an Ornstein-Uhlenbeck process~\cite{SMCrispin}. In order to do this, we begin by expanding the positive-P variables into their steady-state expectation values plus delta-correlated Gaussian fluctuation terms, e.g.
\begin{equation}
\alpha_{ss} \rightarrow \langle\hat{a}\rangle_{ss}+\delta\alpha.
\label{eq:fluctuate}
\end{equation}
Given that we can calculate the $\langle\hat{a}_{i}\rangle_{ss}$, we may then write the equations of motion for the fluctuation terms. The resulting equations are written for the vector of fluctuation terms as
\begin{equation}
\frac{d}{dt}\delta\vec{\alpha} = -A\delta\vec{\alpha}+Bd\vec{W},
\label{eq:OEeqn}
\end{equation}
where $A$ is the drift matrix containing the steady-state solutions, $B$ is found from the factorisation of the diffusion matrix of the original Fokker-Planck equation, $D=BB^{T}$, with the steady-state values substituted in, and $d\vec{W}$ is a vector of Wiener increments. As long as the matrix $A$ has no eigenvalues with negative real parts and the steady-state solutions are stationary, this method may be used to calculate the intracavity spectra via
\begin{equation}
S(\omega) = (A+i\omega)^{-1}D(A^{\mbox{\small{T}}}-i\omega)^{-1},
\label{eq:Sout}
\end{equation}
from which the output spectra are calculated using the standard input-output relations~\cite{mjc}.

In this case
\begin{equation}
A =
\begin{bmatrix}
\gamma_{1} & -\kappa_{1}\alpha_{2} & -\kappa_{1}\alpha_{1}^{\ast} & 0 & 0 & 0 \\
-\kappa_{1}\alpha_{2}^{\ast} & \gamma_{1} & 0 & -\kappa_{1}\alpha_{1} & 0 & 0 \\
\kappa_{1}\alpha_{1} & 0 & \gamma_{2} & -\kappa_{2}\alpha_{3} & -\kappa_{2}\alpha_{2}^{\ast} & 0 \\
0 & \kappa_{1}\alpha_{1}^{\ast} & -\kappa_{2}\alpha_{3}^{\ast} & \gamma_{2} & 0 & -\kappa_{2}\alpha_{2} \\
0 & 0 & \kappa_{2}\alpha_{2} & 0 & \gamma_{3} & 0 \\
0 & 0 & 0 & \kappa_{2}\alpha_{2}^{\ast} & 0 & \gamma_{3}
\end{bmatrix},
\label{eq:Amat}
\end{equation}
and $D$ is a $6\times 6 $ matrix with $\left[\kappa_{1}\alpha_{2},\kappa_{1}\alpha_{2}^{\ast},\kappa_{2}\alpha_{3},\kappa_{2}\alpha_{3}^{\ast},0,0\right]$ on the diagonal. In the above, the $\alpha_{j}$ should be read as their steady-state values.
Because we have set $\gamma_{1}=1$, the frequency $\omega$ is in units of $\gamma_{1}$. $S(\omega)$ then gives us products such as $\delta\alpha_{i}\delta\alpha_{j}$ and  $\delta\alpha_{i}^{\ast}\delta\alpha_{j}^{\ast}$, from which we obtain the output variances and covariances for modes $i$ and $j$ as
\begin{equation}
S^{out}(X_{i},X_{j}) = \delta_{ij}+\sqrt{\gamma_{i}\gamma_{j}} \left(S_{ij}+S_{ji}\right).
\label{eq:Sout}
\end{equation}

\section{Results}
\label{sec:results}

\begin{figure}[tbh]
\centering
\includegraphics[width=0.7\columnwidth]{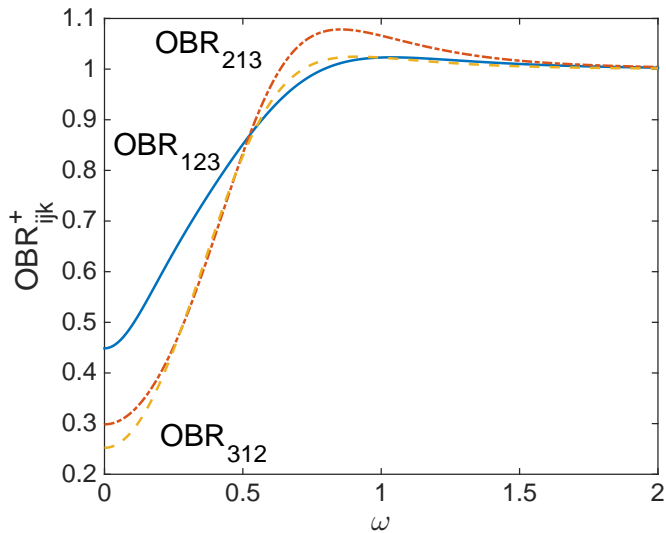}
\caption{(colour online) The three OBR correlations for $\kappa_{1}=5\times 10^{-3}$, $\kappa_{2}=4\kappa_{1}$, $\epsilon=105$, $\gamma_{1}=1$, and $\gamma_{2}=\gamma_{3}=\gamma_{1}/2$. The frequency, $\omega$, is in units of $\gamma_{1}$. OBR$_{123}$ is the solid line, OBR$_{213}$ is the dash-dotted line, and OBR$_{312}$ is the dashed line. All spectra plotted are symmetric about $\omega=0$ and are dimensionless.}
\label{fig:OBRconfig1}
\end{figure}

This system has a very rich parameter regime, with $\kappa_{1}$, $\kappa_{2}$, $\epsilon$, $\gamma_{2}$ and $\gamma_{3}$ all capable of changing independently within any physical constraints. We have performed extensive numerical experiments and present the results for two representative regimes. The first was found to maximise violations of bipartite correlations and give asymmetric steering~\cite{4HGbi}, while the second is interesting because of the different predictions of the various correlations. 

The first parameter set we present is that used previously for bipartite correlations~\cite{4HGbi}, with $\kappa_{1}=5\times 10^{-3}$, $\kappa_{2}=4\kappa_{1}$, $\epsilon=105$, $\gamma_{1}=1$ and $\gamma_{2}=\gamma_{3}=\gamma_{1}/2$, which was shown to give bipartite steering, both symmetric and asymmetric, in all bipartitions. The intention of this parameter set, with lower loss rates at the higher frequencies, is to give the two higher frequency fields more time to interact within the cavity. We find, as shown in Fig.~\ref{fig:OBRconfig1}, that all three possible pairs can steer the remaining mode, but according to the criteria of He and Reid, since the minimum of the sum of the three is $1.44$, genuine tripartite steering is not present. We note that the vLF inequalities were not violated and that since steering is a strict subclass of entanglement, these have missed tripartite entanglement that is present. The better sensitivity of EPR type measures for detecting entanglement has previously been found with bipartite systems~\cite{NGBHcav2}, where the Reid EPR correlations~\cite{MDREPR} have detected entanglement missed by the Duan-Simon positive partial transpose measure~\cite{Duan,Simon}.

\begin{figure}[tbh]
\centering
\includegraphics[width=0.7\columnwidth]{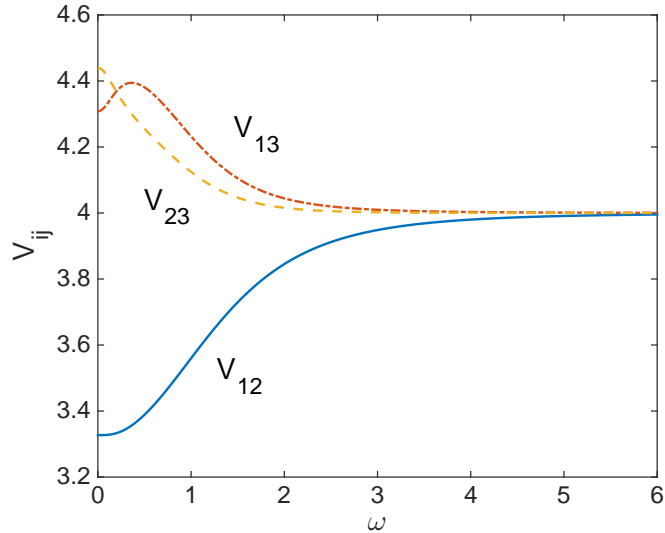}
\caption{(colour online) The three V$_{ij}$ correlations for $\kappa_{1}=10^{-2}$, $\kappa_{2}=0.5\kappa_{1}$, $\epsilon=105$, $\gamma_{1}=1$, $\gamma_{2}=2\gamma_{1}$, and $\gamma_{3}=\gamma_{1}/4$.}
\label{fig:Vijconfig2}
\end{figure}

\begin{figure}[tbh]
\centering
\includegraphics[width=0.7\columnwidth]{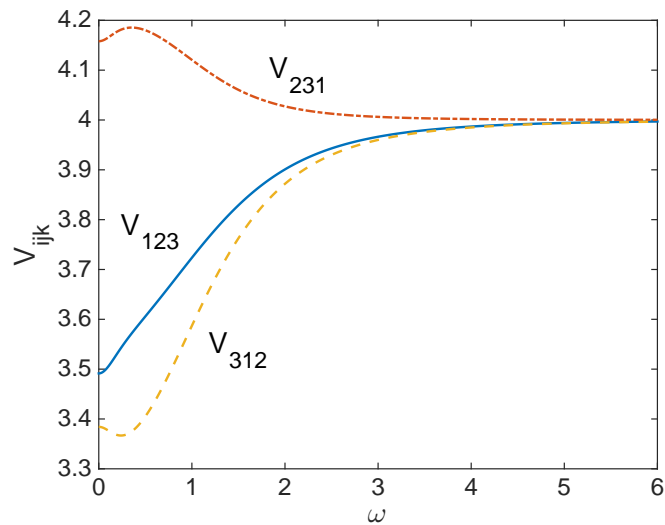}
\caption{(colour online) The three V$_{ijk}$ correlations for $\kappa_{1}=10^{-2}$, $\kappa_{2}=0.5\kappa_{1}$, $\epsilon=105$, $\gamma_{1}=1$ and $\gamma_{2}=2\gamma_{1}$ and $\gamma_{3}=\gamma_{1}/4$.}
\label{fig:Vijkconfig2}
\end{figure}

In the second parameter regime, $\kappa_{1}=10^{-2}$, $\kappa_{2}=0.5\kappa_{1}$, $\epsilon=105$, $\gamma_{1}=1$ and $\gamma_{2}=2\gamma_{1}$ and $\gamma_{3}=\gamma_{1}/4$. We see that only one of the V$_{ij}$ correlations shown in Fig.~\ref{fig:Vijconfig2}, V$_{12}$, drops below 4. This result on its own could be taken to indicate that the system is not tripartite inseparable, but this is not the case. Two of the V$_{ijk}$ shown in Fig.~\ref{fig:Vijkconfig2} drop below a value of 4. Since only one of the possible three violating the inequality is sufficient to prove inseparability, this result is more than sufficient. It is not, however, sufficient to demonstrate genuine tripartite entanglement, since neither of the two $V_{ijk}$ which violate the vLF inequality exhibit values of less than 2.

\begin{figure}[tbh]
\centering
\includegraphics[width=0.7\columnwidth]{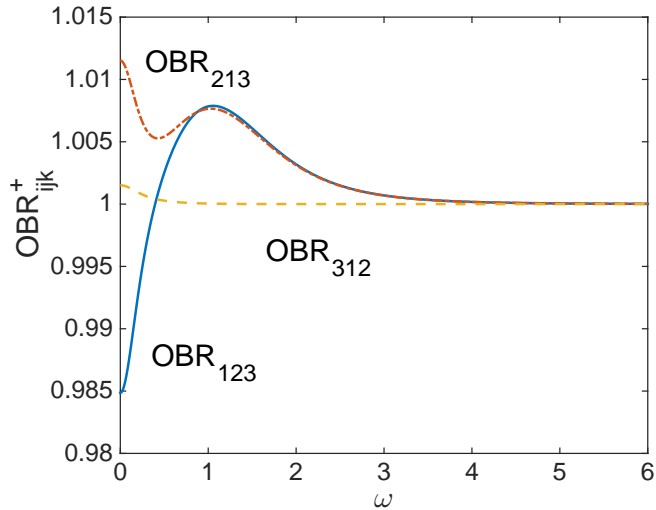}
\caption{(colour online) The three OBR$_{ijk}$ correlations for $\kappa_{1}=10^{-2}$, $\kappa_{2}=0.5\kappa_{1}$, $\epsilon=105$, $\gamma_{1}=1$ and $\gamma_{2}=2\gamma_{1}$ and $\gamma_{3}=\gamma_{1}/4$.}
\label{fig:OBRconfig2}
\end{figure}

For this parameter set we find that the vLF correlations are more efficient at finding separability than the OBR$_{ijk}$, which are shown in Fig.~\ref{fig:OBRconfig2}. We see that only OBR$_{123}$ drops below one, and then by an insignificant amount. This means that, while the participants receiving modes 2 and 3 can combine to steer mode 1 in a marginal fashion which would quite possibly be destroyed by experimental noise, the other two pairings cannot perform steering in any fashion at all by way of Gaussian measurements. Whether steering via non-Gaussian measurements is possible is outside the scope of this article.

\section{Conclusions}
\label{sec:conclusions}

In conclusion, we have analysed a system of cascaded intracavity harmonic generation in terms of tripartite correlations for the detection of inseparability, entanglement, and EPR steering. We have examined two different parameter regimes and found non-classical quantum correlations across two octaves of frequency difference in both of these. In the first regime, EPR like correlations were found to be the best indicator of inseparability and entanglement. In the second, one set of the vLF correlations indicated tripartite inseparability which was missed by the other set, while the EPR correlations were inconclusive, finding only marginal steering in one of the three partitions. Our system, which could have possible applications in multiplexing, is a good physical example of the difficulty of finding versatile measures for tripartite entanglement in mixed systems, with different correlation measures being efficient in different regimes.   

\section*{Acknowledgments}

I would like to thank Margaret Reid for the invitation to contribute to this special issue.

\end{document}